\documentclass[conference]{IEEEtran}
\IEEEoverridecommandlockouts

\usepackage{amsmath,amssymb,amsfonts}
\usepackage{algorithmic}
\usepackage{graphicx}
\usepackage{textcomp}
\usepackage{bbm}

\usepackage[nocomma]{optidef}
\usepackage{xcolor}
\usepackage{cite}
\usepackage{siunitx}
\usepackage{balance}

\def\BibTeX{{\rm B\kern-.05em{\sc i\kern-.025em b}\kern-.08em
    T\kern-.1667em\lower.7ex\hbox{E}\kern-.125emX}}

\newcommand{\eqorn}[1][]{\tag{\theequation\ensuremath{#1}}}
\begin{document}
\bstctlcite{IEEEexample:BSTcontrol}

\title{Enabling Edge Artificial Intelligence via \\ Goal-oriented Deep Neural Network Splitting
\thanks{This work was supported by MIUR under
the PRIN Liquid Edge Contract, by the ANR under the France 2030 program, grant "NF-NAI: ANR-22-PEFT-0003", by the European Union under the Italian National Recovery and Resilience Plan (NRRP) of NextGenerationEU, partnership on “Telecommunications of the Future” (PE00000001 - program “RESTART”).}
}

\author{Francesco Binucci$^{1,2}$, Mattia Merluzzi$^3$, Paolo Banelli$^1$, Emilio Calvanese Strinati$^3$, Paolo Di Lorenzo$^{2,4}$\\
$^1$Department of Engineering, University of Perugia, Via G. Duranti 93, 06128, Perugia, Italy\\
$^2$Consorzio Nazionale Interuniversitario per le Telecomunicazioni, Viale G.P. Usberti, 181/A, 43124, Parma, Italy\\
$^3$CEA-Leti, Université Grenoble Alpes, F-38000 Grenoble, France\\
$^4$DIET department, Sapienza University of Rome, via Eudossiana 18, 00184, Rome, Italy}

\maketitle

\begin{abstract}
Deep Neural Network (DNN) splitting is one of the key enablers of \textit{edge} Artificial Intelligence (AI), as it allows end users to pre-process data and offload part of the computational burden to nearby Edge Cloud Servers (ECSs). This opens new opportunities and degrees of freedom in balancing energy consumption, delay, accuracy, privacy, and other trustworthiness metrics. In this work, we explore the opportunity of DNN splitting at the edge of 6G wireless networks to enable low energy cooperative inference with target delay and accuracy with a \textit{goal-oriented} perspective.
Going beyond the current literature, we explore new trade-offs that take into account the accuracy degradation as a function of the Splitting Point (SP) selection and wireless channel conditions. Then, we propose an algorithm that dynamically controls SP selection, local computing resources, uplink transmit power and bandwidth allocation, in a goal-oriented fashion, to meet a target goal-effectiveness.
To the best of our knowledge, this is the first work proposing adaptive SP selection on the basis of all learning performance (i.e., energy, delay, accuracy), with the aim of guaranteeing the accomplishment of a \textit{goal} (e.g., minimize the energy consumption under latency and accuracy constraints). 
Numerical results show the advantages of the proposed SP selection and resource allocation, to enable energy frugal and effective edge AI.


\end{abstract}


\section{Introduction}
6G will offer the fusion of cyber physical spaces, mobile connect-compute networks, Artificial Intelligence (AI) and Machine Learning (ML). 
These new services will bond communication and computing for a continuous data distillation, processing and exchange among heterogeneous intelligent agents and network infrastructure resources \cite{calvanese_strinati_6g_2019},\cite{jiang2021road} that will be flexibly accessed on demand. Also, they must be enabled with stringent constraints in terms of energy consumption, latency, reliability and trustworthiness among others. To this end, Multi-access Edge Computing (MEC) moves computational and storage resources at the edge of wireless networks, thus keeping computation as close as possible to the data source. Among several services, we focus on Artificial Intelligence (AI) at the edge, whose goal is to enable AI algorithms (learning and inference) in a cooperative and distributed way, involving end devices, MEC resources, and central clouds \cite{Merluzzi2023_access}. In this paper we focus on the inference phase, i.e., edge inference (EI). 
The latter opens new challenges to the orchestration of the network resources, which include an optimal choice for \textit{i)} splitting computations between local devices and remote ones, \textit{ ii)}  data compression, and \textit{iii)} communication design (e.g., bandwidth, transmit power, interference management, etc.). For instance, an interesting goal is to reduce the computational burden at the local edge devices (EDs), by optimally providing enough MEC resources to perform a task, with a prescribed level of \textit{reliability}, and within predefined latency limits. In this context, \textit{goal-oriented} communications (GOCs) represent an emerging topic \cite{calvanese_strinati_6g_2019, chaccour2022less, shao2021learning, strinati20216g, kountouris2021semantics,dilo_GO_2023} that will help mitigating the unsustainable increase of data traffic, by transmitting the minimum amount of information necessary to perform a specific task with target effectiveness \cite{MerluzziGO2023} (i.e., to achieve a specific \textit{goal}). The enablers of GOCs span from physical layer aspects up to the application ones, with a cross-layer perspective where bits, information distortion, and application performance are 
\textit{jointly} optimized, in order to strike the best trade-offs among energy consumption, latency, and learning performance.
This paper proposes a goal-oriented computation split between EDs and Edge Cloud Servers (ECSs), building on the concept of Deep Neural Network (DNN) splitting \cite{Matsubara2023}. Specifically, splitting a DNN means to separate it into a head and tail structure, to perform part of the computation at an ED (e.g., the head at the device collecting data), and the remaining portion at another node (e.g., the tail at an ECS). This can be seen as a cooperation problem, where the costs (i.e., latency and energy expenditure due to the computation/transmission) are broken down between the head and the tail focusing, for instance, on the energy/latency trade-offs, on the quality of the wireless channels \cite{cnn_splitting}, or on the target \textit{goal-effectiveness}.

\textbf{Related works.} The benefits of DNN splitting strategies in MEC environments have been already investigated in some recent works. The authors of \cite{hu2019dynamic} and \cite{manasi2020neupart} consider DNN partitioning schemes to minimize the latency and the user equipment energy consumption, respectively.
\begin{figure*}[t]    \centering
\includegraphics[width=.6\textwidth]{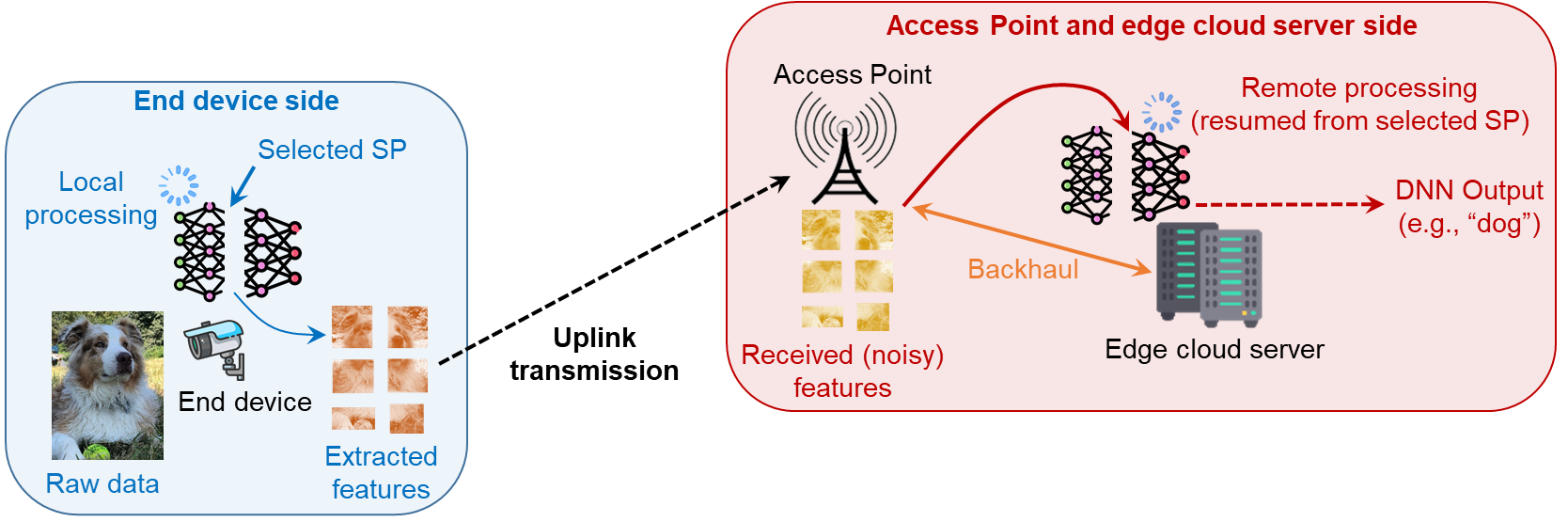}
    \caption{Illustration of the considered DNN splitting scenario.}
    \label{fig:system_model}
    \vspace{-0.4 cm}
\end{figure*}
Some works, like \cite{shao2021learning}, witness how goal-oriented compression schemes can be seen as a specific case of DNN splitting. In these scenarios, the EDs run an \textit{encoder} network, while the ECS employs a \textit{decoder} network. The encoders compress data taking into account the informativeness with respect to the output of a specific inference task, retrieved by the decoder. Along this line, in \cite{binucci2023multi,eusipco2023}, we propose an inference architecture where a convolutional neural network is split between a Convolutional Encoder (CE), used to perform a goal-oriented compression at the device-side, and a Convolutional Classifier (CC) to complete the inference task at the ECS side. Although these works consider the dynamic optimization of the transmission and computational resources, they do not address the problem of optimally and dynamically selecting the splitting point (SP) of the employed architectures. Differently, the work in \cite{cnn_splitting}, proposes a dynamic SP selection strategy that reduces the energy consumption of a user equipment to perform an edge-assisted image classification task with target delay constraints. However, \cite{cnn_splitting} does not consider the performance degradation (e.g., the effects on the accuracy) introduced by the noisy wireless transmission, and its dependency on the SP selection.

\textbf{Our contribution.} In this work, we focus on \textit{goal-oriented DNN splitting} to enable low energy and timely cooperative edge inference with target delay and accuracy performance, dynamically and jointly optimizing the SP selection and the network resources. Specifically, focusing on an image classification task, we first evaluate the impact of noise injection at different SPs during the inference phase. Then, taking a step forward from our previous work \cite{cnn_splitting}, we incorporate this knowledge in an optimal resource allocation strategy, which dynamically selects also the best SP, with the goal of minimizing the ED energy consumption under latency and accuracy constraints. To the best of our knowledge, this is the first work that takes into account the accuracy degradation due to noisy transmission of intermediate extracted features, depending on the SP selection and wireless channel conditions. The proposed algorithmic framework hinges on Lyapunov stochastic optimization, which enables efficient adaptive solutions, without needing any prior knowledge of the statistics of the wireless channels and the task arrival process. Finally, numerical results assess the performance of our method, illustrating its benefits in terms of energy-accuracy trade-off on realistic image classification tasks. 
\section{System model}

Let us consider the system model illustrated in  Fig. \ref{fig:system_model},  composed by a single ED (e.g., a camera) equipped with computing and communication resources, and an Access Point (AP) connected to an ECS through an ideal backhaul. Both the ED and the ECS are embarked with the same Deep Neural Network (DNN) model, which is assumed to be trained and ready to treat inference requests on the data collected by the ED.
In a slotted fashion, the ED generates a batch $\mathcal{B}$ of $card(\mathcal{B})$ new patterns to be inferred (e.g., a set of images). The inference can partly take place at the ED and at the ECS, thanks to DNN splitting. In particular, for each request, the ED selects an SP based on the current status of communication and computation resources (availability, quality of wireless channels, etc.). The choice of the SP affects the size (dimensionality) of the data to be transmitted and the portion of computations to be performed locally, thus impacting the energy consumption of the ED to process and transmit, as well as the delay. Therefore, once an SP $k$ is selected, the ED processes data up to SP $k$, and transmits the extracted features to the AP. Then, the AP transfers the features to the ECS, which resumes the computations up to the last layer of the DNN, to output the inference result (e.g., a set of labels). The set of SPs is denoted by $\mathcal{J}=\{0,1,\ldots,k,\ldots,J\}$, where $0$ denotes the full offloading case (raw data are transmitted), and $J$ the full local computation. Each SP $k$ can be represented by a tuple $\mathbf{s}_k=[L_k,F_k]$, with $L_k$ the number of its output features, and $F_k$ the number of FLOPS needed to process data across layers between SP $k-1$ and SP $k$
\subsection{Local computation model: Delay and energy consumption}
We consider a slotted system indexed by $t=1,2,\ldots,$. At the beginning of each slot, a new request is issued and a control decision is taken on the system resources (e.g., communication and computing). At the ED side, denoting by $f_l(t)$ the CPU clock frequency at time instant $t$, and by $\eta_l$ the number of FLOPS per CPU cycle (hardware dependent), and assuming that the whole batch $\mathcal{B}(t)$ is serially handled through the selected SP $k$, the local computing delay to extract features can be written as follows:
 \begin{equation}\label{local_comp_delay}
     D^{\text{comp}}_l(t)=\begin{cases}
     \displaystyle\frac{card(\mathcal{B}(t))\sum_{j=0}^kF_j}{\eta_l f_l(t)}, \;\text{if}\;k(t)>0\\
    0, \;\text{otherwise}
     \end{cases}
 \end{equation}
Similarly, using the typical cubic model for the CPU dynamic power consumption \cite{energy_processor_article}, the local computing energy consumption is modeled as
\begin{equation}\label{energy_comp}
 E_l^{\text{comp}}=\begin{cases}
 \displaystyle\frac{card(\mathcal{B}(t))\sum_{j=0}^kF_j\kappa f_l^2(t)}{\eta_l},\;\text{if}\;k(t)>0 \\
     0, \;\text{otherwise},
 \end{cases}
\end{equation}
with $\kappa$ the effective switched capacitance of the processor \cite{energy_processor_article}.
\subsection{Transmission model: Delay and energy consumption}
In this work, we assume an analog goal-oriented transmission \cite{eusipco2023}. The impact of DNN splitting with digital modulations is left for future work, although analog modulations are becoming popular in the context of edge learning problem with the concept of over-the-air computation for, e.g., federated learning \cite{Hellstrom2021}. Therefore, we assume a base-band PAM signal expressed as follows \cite{goldsmith2005wireless}: 
\begin{equation}\label{eq:pam_wave}
    u(\tau)=\sum\nolimits_{k=-\infty}^{+\infty}a_{k}p_{g}(\tau-kT_{s}),
\end{equation}
where $a_{k} \in \mathbb{C}$ is the $k$-th transmitted symbol, $p_{g}(\tau)$ is the pulse shaper, and $S_{r}={1}/{T_{s}}$ is the symbol rate. 
We consider a Squared Root Raised Cosine (SRRC) shaping pulse $p_{g}(\tau)$, with two-sided baseband noise-equivalent bandwidth $W=S_r$, roll-off factor $\beta=0.25$ \cite{goldsmith2005wireless}, and unit energy. This way, we can assume an Inter-Symbol Interference (ISI)-free system.  
We consider analog modulations, where the outputs $z_{k}$ of a selected SP $k$ are multiplexed on the I-Q components, and \emph{directly} mapped into the (analog) QAM symbols by $a_{k}\!=\!z_{2k-1}\!+\!j z_{2k}$. The transmitted signal $u(\tau)$ is received through a Single-Output (SISO) noisy flat-fading wireless channel $h(t)\in\mathbb{C}$ and, after matched-filtering and one-tap equalization, it is passed to the ECS for classification. 
Then, the instantaneous SNR at time instant $t$ is
\begin{equation}\label{eq:snr_equation}
    \gamma(t)=\frac{p_{\text{tx}}(t)|h(t)|^2}{W(t)N_{0}},
\end{equation}
where $N_{0}$ is the noise power spectral density (PSD), $W(t)$ is  the bandwidth, and $p_{\text{tx}}(t)$ is the transmit power. In each slot $t$, we optimize the transmit power, the bandwidth, and the SP. For a generic SP $k$, the overall communication delay can be written as follows
\begin{equation}\label{tx_delay}
     D^{\text{tx}}(t)=\begin{cases}\displaystyle\frac{(1+\beta)L_kcard(\mathcal{B}(t))}{2W(t)},\;\text{if}\;k(t)<J\\
     0,\;\text{otherwise}.
     \end{cases}
\end{equation}
Then, the transmit energy consumption is
\begin{equation}\label{tx_energy}
    E^{\text{tx}}(t)=p_{\text{tx}}(t)D^{\text{tx}}(t).
\end{equation}
\subsection{Remote computation delay}
Once the data related to SP $k$ are received by the AP and transferred to the ECS, the computation is resumed from $k$ up to the output layer of the DNN. Then, denoting by $f_r(t)$ the CPU cycle frequency of the ECS at time $t$, and by $\eta_r$ the number of FLOPS per CPU cycle, the remote computation delay can be written as:
\begin{equation} D^{\text{comp}}_r=\begin{cases}\displaystyle\frac{\left(\sum_{j=0}^JF_j-\sum_{j=0}^k F_j\right)card(\mathcal{B}(t))}{\eta_r f_r(t)},\;\text{if}\;k(t)<J\nonumber\\
     0,\;\text{otherwise},    
     \end{cases}
 \end{equation}
In this paper, as in \cite{cnn_splitting}, we assume that $f_r(t)$ cannot be optimized, but it is an exogenous variable that depends on the time-varying availability of the ECS.
Finally, the end-to-end delay can be written as the sum of all components 
\begin{equation}\label{e2e_delay}
    D^{\text{tot}}(t)=D_l^{\text{comp}}(t)+D^{\text{tx}}(t)+D_r^{\text{comp}}(t),
\end{equation}
while the overall ED energy consumption reads as
\begin{equation}\label{energy_tot}
    E^{\text{tot}}(t)=E_l^{\text{comp}}(t)+E^{\text{tx}}(t)
\end{equation}

\subsection{Inference accuracy: The effect of noise and SP selection}
\begin{figure}[t]
    \centering
    \includegraphics[width=1.00\linewidth]{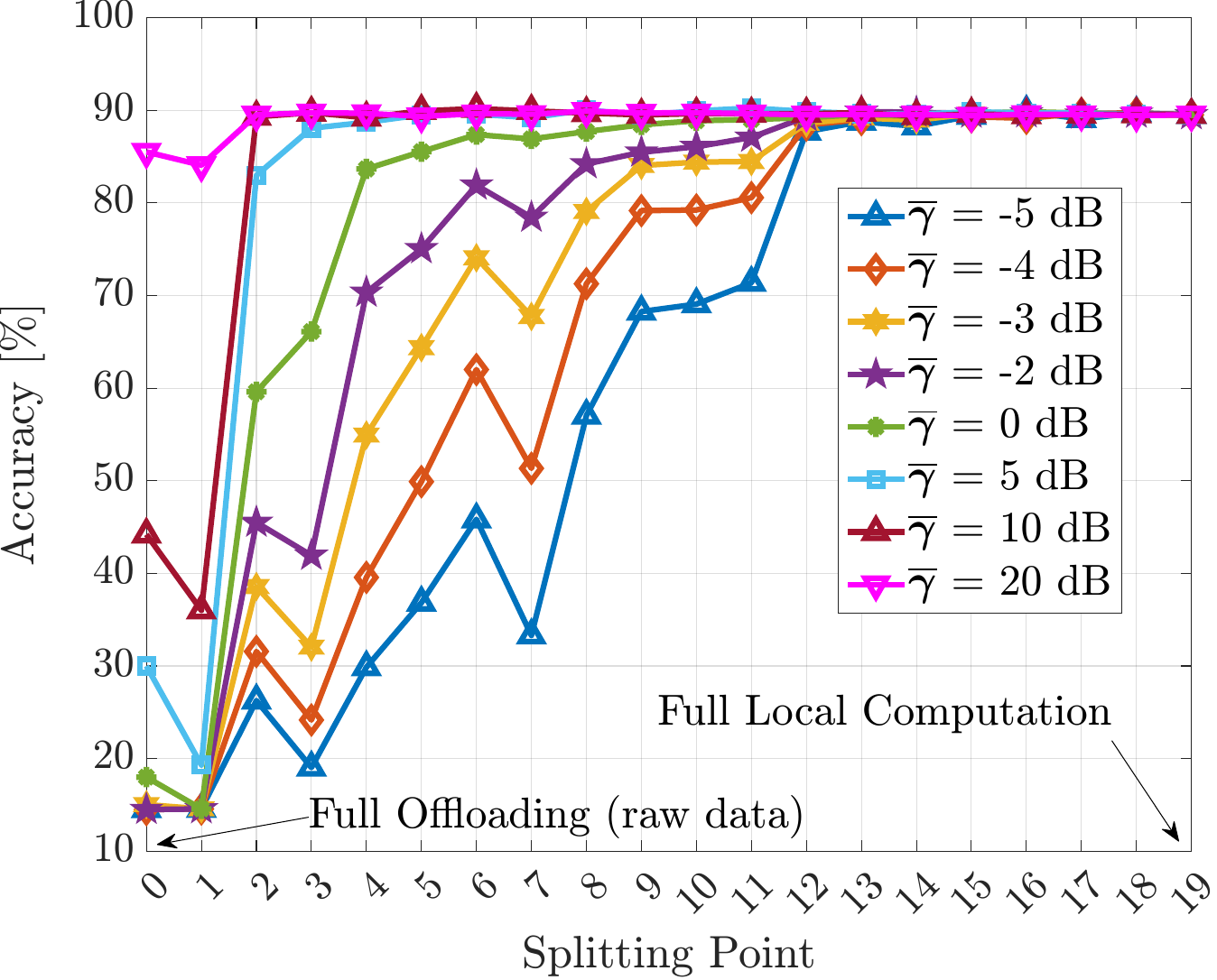}
    \caption{Accuracy vs Splitting Point for different average SNRs $\overline{\gamma}$.}
    \label{fig:accuracy_vs_sp}
\end{figure}

In this section, we evaluate the accuracy of a DNN as a function of the splitting point at different SNRs during wireless communication. As in \cite{cnn_splitting}, we employ a pre-trained version of the \textit{MobileNetV2} network\cite{sandler2018mobilenetv2}, which is composed of 20 possible SPs, including the full offloading and the full local computation case. 
We perform an image classification task based on the \textit{Intel-Scene classification data-set}\cite{bansal2019intel}, containing 17000 RGB images of landscapes, belonging to 6 different classes. We include $14000$ samples in the training-set and $3000$ samples in the test-set. The images have been resized to $224 \times 224$ pixels, in order to match the input-size of the classification network, and the last layer has been resized to meet the number of possible classes contained in the data-set. The model was fine-tuned by training it for 30 epochs, considering \textit{Adam} as optimizer with a learning-rate $lr=\num{1e-3}$. As usual in classification tasks, we employ the well-known \textit{cross-entropy} loss as objective function. Fig. \ref{fig:accuracy_vs_sp} shows the accuracy in the inference phase adding white Gaussian Noise at different SPs, to mimic different SNR conditions. Overall, we can notice that the outputs of the top-layers (i.e., the ones deeper in the architecture) are less affected by the noise with respect to the output of the bottom-layers.  Indeed, according to \cite{he2023law}, the discrimination ability of DNNs tends to grow as we go deeper in the network architecture. Thus, the effect of the noise is mitigated in the final layers of the network. However, we can also notice that some SPs (e.g., 3, 7) may exhibit an irregular behaviour. Indeed, as reported in \cite{he2023law}, for some network architectures (e.g., residual networks) the data-separation law could present irregular behaviours, although representing a valid approximation.
Fig. \ref{fig:accuracy_vs_sp} suggests that, under noisy wireless communication, the SP selection should take into account energy, delay and accuracy, and not only energy and delay as in \cite{cnn_splitting}. Therefore, in the next section, we focus on devising an algorithm that jointly selects SP, as well as wireless and computing resource, to balance energy, delay, and accuracy.

\section{Problem formulation and solution}
Due to the energy resource scarcity typical of EDs, the aim of this work is to minimize the ED energy consumption in \eqref{energy_tot} under average delay constraints as in \cite{cnn_splitting}, and a target accuracy constraint. A long-term problem is formulated: 
\begin{mini!}|s|[2]
{\Phi(t)}{\lim_{T\to\infty} \frac{1}{T} \sum\nolimits_{t=1}^{T} \mathbb{E}\{E^{\text{tot}}(t)\}\eqorn\label{eq:long_term}}{}{}
\addConstraint{(a)\;}{\lim_{T\to\infty} \frac{1}{T} \sum\nolimits_{t=1}^{T} \mathbb{E}\{D^{\text{tot}}(t)\}\leq D^{\text{avg}}}{}{}\nonumber
\addConstraint{(b)\;}{\lim_{T\to\infty} \frac{1}{T} \sum\nolimits_{t=1}^{T} \mathbb{E}\{G(k(t), \gamma(t)\}\geq G^{\text{avg}}}{\hspace{0.5cm}}\nonumber
\addConstraint{(c)\;}{p_{\text{tx}}(t) := \frac{\gamma(t) N_{0}W(t)}{|h(t)|^2} \leq p_{\text{tx}}^{\max}},\quad \forall t  \nonumber
\addConstraint{(d)\;}{0 \leq W(t) \leq W_{\max}}\cdot\mathbbm{1}(k(t)<J),\quad \forall t   \nonumber
\addConstraint{(e)\;}{k(t) \in \mathcal{J}, \hspace{6pt} \gamma(t) \in \mathbb{R}_0^+},\quad \forall t  \hspace{6pt} \nonumber
\addConstraint{(f)\;}{f_{l}^{\min}\cdot\mathbbm{1}(k(t)\!>\!0)\! \leq\! f_l(t)\!\leq\! f_{l}^{\max}\cdot\mathbbm{1}(k(t)\!>\!0),\, \forall t  }{}{}\nonumber
\end{mini!}
where $\Phi(t)=[W(t), k(t), \gamma(t),f_l(t)]$ is the set of the optimization variables. $(a)$ is the \textit{long-term} latency constraints, while $(b)$ is the accuracy constraint; $(c)$ and $(d)$ encompass the constraints on the transmission power and bandwidth, respectively, while $(e)$ imposes the SP $k(t)$ to be chosen from the set $\mathcal{J}$ of all possible SPs and the SNR constraint; finally, $(f)$ imposes the CPU cycle frequency of the ED to be chosen between a minimum and a maximum value, while being $0$ in case of full offloading decision. 

To solve problem, we build on Lyapunov stochastic optimization. First, we define two \textit{virtual queues} to handle the long-term constraints $(a)$-$(b)$. The virtual queues evolve as:
\begin{equation}
    \begin{split}     (a)\;Z(t+1)&=\max(0,Z(t)+\mu(D^{\text{tot}}(t)-D^{\text{avg}}))\\
    (b)\;Y(t+1)&=\max(0,Y(t)+\lambda(G^{\text{avg}}-G(t))),
    \end{split}
\end{equation}
where $\mu$, and $\lambda$ are positive step-sizes. Then we build the Lyapunov function and the associated Lyapunov Drift plus penalty (DPP) function
\begin{equation}\label{dpp}
\begin{split}
    L(\Theta(t))&=Z^2(t)+Y^2(t)\\
    \Delta_{p}(\Theta(t))&=\Delta(\Theta(t))+V\mathbb{E}\{E^{\text{tot}}(t)|\Theta(t)\},
\end{split}
\end{equation}
where $\Theta(t)=\{Z(t),Y(t)\}$, and $\Delta(\Theta(t))=\mathbb{E}\{L(t+1)-L(t)|\Theta(t)\}$ is the conditional Lyapunov Drift \cite{neely2010stochastic}, while the parameter $V$ controls the trade-off between objective function minimization and virtual queues stability. 
Removing expectations and exploiting upper-an upper bound of the DPP \cite{neely2010stochastic} that is omitted  due the lack of space, the optimization  decouples among different slots, with virtual queues representing dynamic penalties that drive the system towards constraint guarantees. In particular, the mean rate stability of each virtual queue\footnote{For a virtual queue $G$ is defined as $\lim_{T\to\infty}\frac{\mathbb{E}\{G(T)\}}{T}=0$.} guarantees the associated constraint \cite{neely2010stochastic}. At each slot, based on the observation of virtual queues and random context parameters, we solve the following problem (cf. \eqref{e2e_delay}, \eqref{energy_tot}):
\begin{align}\label{slot_prob}
    &\underset{\Phi(t)}{\min}\;\; \mu Z(t)D^{\text{tot}}(t)-\lambda Y(t) G(k(t),\gamma(t))+VE^{\text{tot}}(t)\\
    &\qquad\text{s.t. $(c)$-$(f)$ of \eqref{eq:long_term}}\nonumber
\end{align}
Problem \eqref{slot_prob} is a mixed integer convex program. In particular, variables $\gamma(t)$ and $k(t)$ are integer, while the problem is convex in $W(t)$ and $f_l(t)$. Also, once the SP is fixed, the problem can be split into a communication and a computation sub-problem.
\subsection{Communication sub-problem solution}
Since the main goal of this work is to show the importance of a goal-oriented SP selection, we perform an exhaustive search over $\gamma(t)$ and $k(t)$, while we reserve less complex optimization strategies to future work, fundamental in multiple users scenarios. Once $\gamma(t)$ and $k(t)$ are fixed, the problem is convex in $W(t)$, which can be easily computed in closed form:
\begin{equation}
    W^*(\gamma,k)=\min\left(\frac{p_{\text{tx}}^{\max}|h(t)|^2}{\gamma(t)N_0},W_{\max}\right)\cdot\mathbbm{1}(k(t)<J),
\end{equation}
i.e., we always transmit with the maximum available bandwidth, since the objective function is 
monotonically decreasing with respect to $W$.
\subsection{Computation sub-problem solution}
The optimal ED clock frequency can be easily computed by nulling the derivative of the objective function and imposing the KKT conditions \cite{boyd2004convex}, thus ending up the following closed form expression
\begin{equation}
    f_l^*(t, k(t))=\left[\sqrt[3]{\frac{\mu Z(t)}{2\kappa V}}\right]_{f_{l}^{\min}\cdot\mathbbm{1}(k(t)>0)}^{f_{l}^{\max}\cdot\mathbbm{1}(k(t)>0)}
\end{equation}
Finally, given the optimal bandwidth and clock frequency for each pair $(k(t),\gamma(t))$, the optimal solution can be found as
\begin{equation}\label{opt_solution}
    [W^*(t),\gamma^*(t),k^*(t)]= \underset{W^*, f_{l}^*(\gamma,k),\gamma,k}{\text{argmin}}\;\mathcal{O}(t),
\end{equation}
with $\mathcal{O}(t)$ denoting the objective function in \eqref{slot_prob}.

\section{Simulation Results}
In this section, we assess the performance of the developed dynamic resource allocation strategy.
We run the optimization strategies for $N=10^4$ slots, averaging results by removing the transient interval. Similarly as in our previous works \cite{eusipco2023}, we model the accuracy function in \eqref{eq:long_term} $G(k(t),\gamma(t))$ using a Look-UP table (LUT) obtained through some simulations on the test-set of the \textit{Intel-Scene classification data-set}\cite{bansal2019intel}.
 
\textit{\underline{Computing Model:}} the maximum CPU clock frequency of the ED is $f_{l}^{\max}=1.4$ GHz, and a conversion factor $\eta_l=50$ FLOPS per CPU cycle  (cf. \eqref{local_comp_delay}) is assumed. We consider an effective switched capacitance $\kappa=\num{1.097e-27} \left[\frac{s}{\text{cycles}}\right]^3$. Instead, the ECS is equipped with a maximum clock frequency $f_{r}^{\max}=4.5$ GHz and $\eta_r=2000$ FLOPS per CPU cycle. The ECS availability varies across the slots according to $f_{r}(t) = \alpha_{r}(t) f_{r}^{\max}$, where $\alpha_{r}(t) \sim \mathcal{U}(0,1)$. The computational costs associated to the inference network have been computed considering the model in \cite{sandler2018mobilenetv2} and stored in specific LUTs used to drive the optimization strategy. The features dimensionality and the number of FLOPS needed to compute across SPs can be also found in \cite[Fig. $1$]{cnn_splitting}

\textit{\underline{Wireless communication model:}} we consider an ED with a maximum transmit power $p_{\text{tx}}^{\max}=300$ mW, and a bandwidth $W_{\max}=10$ MHz; the noise PSD is set to $N_{0}=-174$ dbm/Hz, with a receiver noise figure $F=5$ dB. The channel is characterized by Rayleigh fading with unit variance, while the path loss is set to different values visible in the illustrated results.
The number $card(\mathcal{B}(t))$ of images generated at each instant is modelled as a Poisson process with $\lambda=5$. In all simulations we set $\gamma(t) \in \{-5, -4, -3, -2, 0, 5, 10, 20\}$ dB.

In the first set of results we highlight the benefits of the dynamic selection of both the Splitting Point $k$ and the SNR $\gamma$, compared to strategies that fix one or the other. We set a latency constraint $D^{\text{avg}}=50$ ms and we select the parameter $V$ (cf. \eqref{dpp}) that minimizes the energy under such constraints, i.e., all comparisons are performed with the same end-to-end delay. We test the allocation strategy considering different accuracy constraints, namely $\mathcal{G}=\{70, 75, 80, 85\}$ \%. We assume different propagation conditions, corresponding to $115, 120, 125$ dB of path loss.

\begin{figure}[ht]
    \centering
\includegraphics[width=1.00\linewidth]{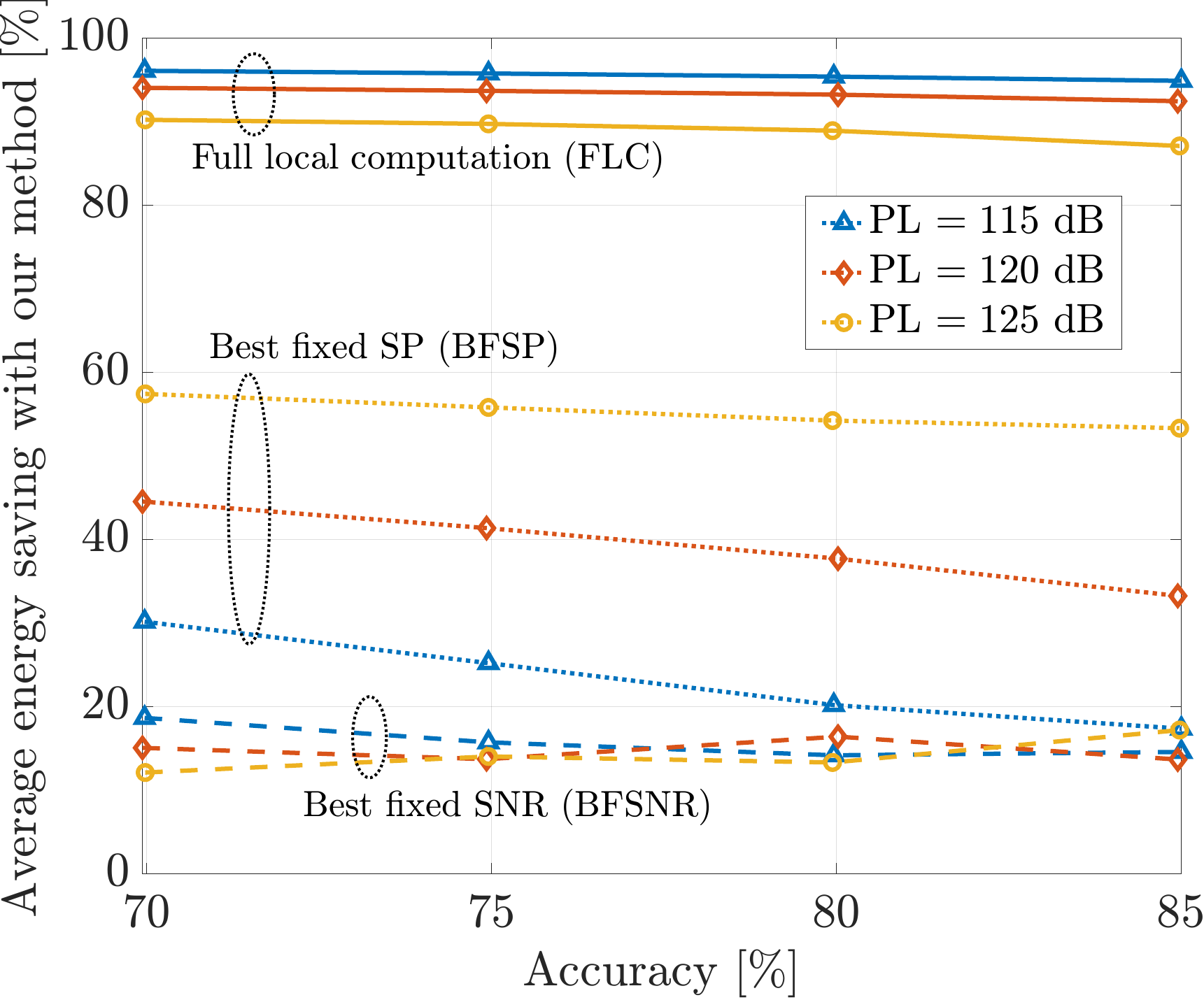}
    \caption{Energy saving of our method with respect to benchmark strategies}
    \label{fig:energy_accuracy_trade_off_fixed_sp}
\end{figure}

In Fig. \ref{fig:energy_accuracy_trade_off_fixed_sp}, we show the energy saving achieved with our method with respect to three benchmarks: \textit{i)} \textit{full local computation (FLC - solid lines):} , i.e. all computations are performed locally at the device \textit{ii)} \textit{Best fixed SP (BFSP - dotted lines):} the SP is fixed and the SNR is optimized vi the adaptive algorithm; to make a fair comparison, we illustrate the results obtained with the SP achieving the lowest energy consumption; \textit{iii)} \textit{Best fixed SNR (BFSNR - dashed lines):} the SNR is fixed and the SP is optimized; again, the SNR achieving the lower energy consumption is considered for fair comparison. 

\textit{\underline{Comparison with \textit{FLC}: }} First, we can notice how our method achieves considerable energy saving in terms of energy consumption with respect to performing all computations locally. The gain obtained by the proposed strategy varies in the range $[88,95]\%$, depending on the path loss and the accuracy constraints. As excepted, the higher the path loss, the lower is the energy saving due to the need of increasing the  transmission energy consumption, but also the fact that more computations need to be performed locally to achieve the target accuracy (cf. Fig. \ref{fig:accuracy_vs_sp}). 

\textit{\underline{Comparison with \textit{BFSP}: }} with respect to the strategy with \textit{BFSP}, the energy saving is still remarkable, ranging from $20\%$ and $60\%$. However, differently from the previous benchmark, it decreases as a function of the path loss. This is due to the reduced degrees of freedom in pushing computations locally or towards the edge. In fact, with the dynamic SP selection strategy (i.e., our method), the transmission cost can be mitigated balancing between local and remote processing in a dynamic and adaptive fashion. 
\begin{figure}
    \centering
    \includegraphics[width=1.00\linewidth]{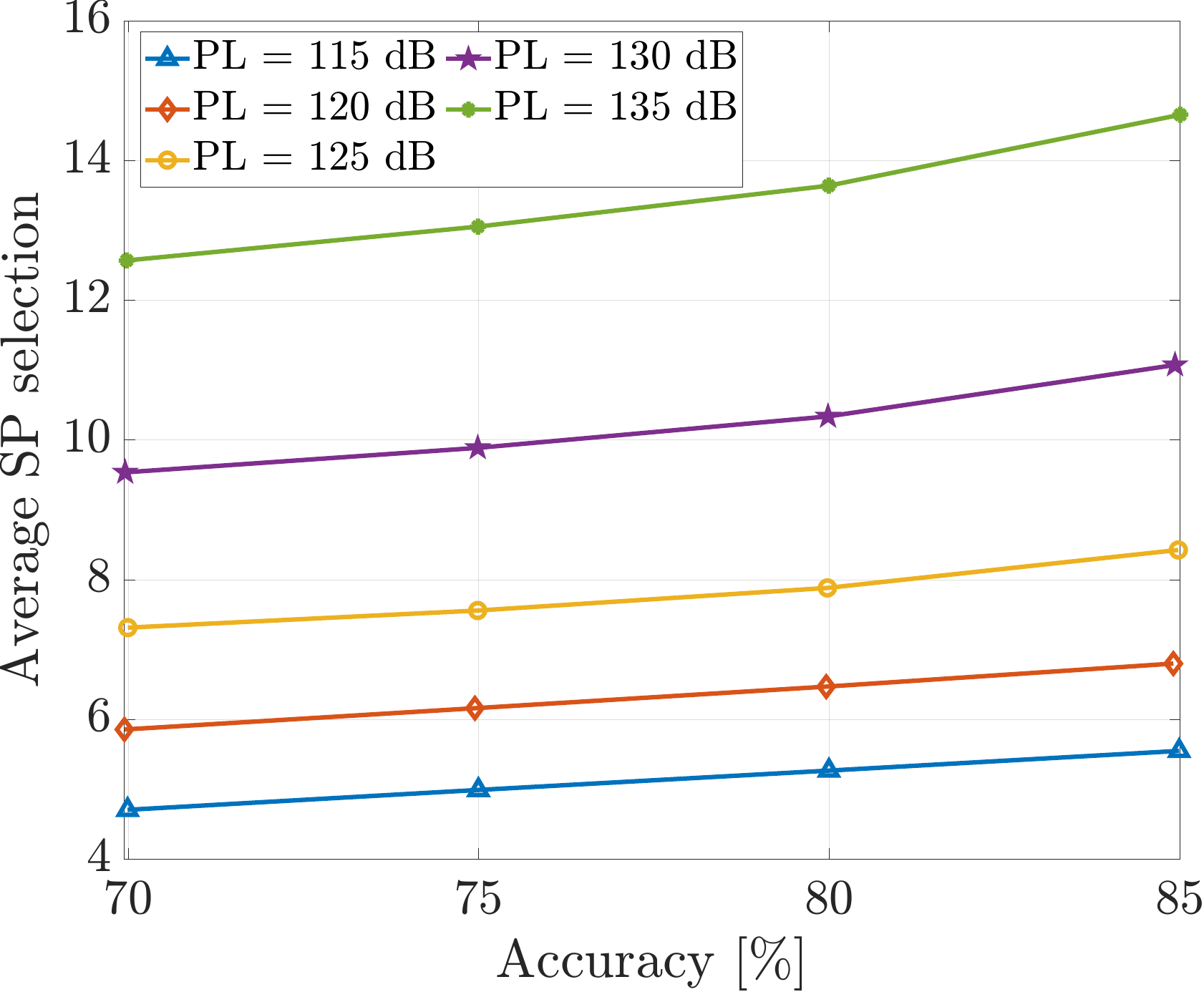}
    \caption{Average splitting point vs Accuracy for different channel conditions.}
    \label{fig:dynamic_accuracy_vs_splitting_point}
\end{figure}

\textit{\underline{Comparisons with \textit{BFSNR}:}} 
we note that our method achieves an appreciable reduction in terms of energy consumption, in the order of $10$-$20\%$. The impact of the path loss on this gain is negligible. Also, the impact of the dynamic selection of the SNR is lower with respect to the dynamic optimization of the Splitting Point (cf dashed Fig. \ref{fig:energy_accuracy_trade_off_fixed_sp}). We point out that, to better evaluate the impact of the dynamic SNR selection, $\gamma(t)$ should be considered as a continuous optimization variable while in this preliminary work we are treating it as discrete. We leave this analysis for future investigations.

\underline{\textit{On the impact of the target accuracy on the SP selection:}} as an additional informative result, in Fig. \ref{fig:dynamic_accuracy_vs_splitting_point} we plot the average SP selection (i.e., a measure of the depth of local computations) obtained by the proposed strategy as a function of the accuracy, for different path loss conditions. As expected, increasing the accuracy constraint forces the system to \textit{autonomously} choose a deeper splitting point (cf. Fig. \ref{fig:accuracy_vs_sp}). Also, the depth of the average splitting point increases as function of the path loss, as already remarked in \cite{cnn_splitting}. Indeed, as the channel conditions worsen, the transmission cost increases, thus encouraging the ED to perform more local processing, of course at the cost on increased energy.  

\textit{\underline{Comparison with accuracy unaware strategies: }} 
finally, to better highlight the benefits of the goal-oriented SP selection, we test the dynamic resource allocation following the same philosophy of \cite{cnn_splitting}, removing the accuracy constraint (cf. $(b)$ in \eqref{eq:long_term}) and focusing only on energy and latency.
\begin{table}[ht]
    \centering
    \caption{Effective accuracy of the accuracy unaware strategy.}
    \begin{tabular}{|c|c|c|c|}
         \hline
         $PL [dB]$ & 115 & 120 & 125\\
         \hline
         $G_{eff} [\%]$ & 17.302 & 22.17 & 33.40\\ 
         \hline
    \end{tabular}
    \label{tab:accuracy_unaware_performance}
\end{table}

As witnessed by Table \ref{tab:accuracy_unaware_performance}, the accuracy-unaware strategy leads to a dramatic reduction in terms of learning performance. With a lower path loss, the system chooses the first SPs of the network, which are also the most affected by the noise. Conversely a higher path loss naturally pushes the strategy to select deeper SPs, with a consequent accuracy increase.

\section{Conclusion and future work}
We proposed a dynamic optimization strategy for goal-oriented DNN splitting with latency and accuracy guarantees. First, we investigated the effect of SP selection and wireless communication noise on the accuracy of a cooperative inference task. Building on the acquired knowledge of this behavior, we tailored a dynamic algorithm that jointly optimizes DNN splitting, communication and computation resources. Numerical results show the effectiveness of the proposed approach to meet the prescribed constraints also in noisy environments, and encourages to further investigate this research line.
Future research directions include the evaluation of the proposed strategy including outage probability constraints,  the joint optimization of the ED and the ECS resources, and the extension to multi-user scenarios should be investigated. Also, early exiting as a way of reducing energy and/or latency is worth being investigated \cite{Bullo2023}. 
\bibliographystyle{IEEEtran}
\bibliography{ICC2024}

\end{document}